 \documentclass[preprint,showkeys,preprintnumbers,amsmath,amssymb]{revtex4}

 \usepackage{graphicx}  

\newcommand{\D}{\partial}
\newcommand{\ve}[1]{{\bf\underline #1}}

\newcommand{\Dt}[1]{\frac {\D #1} {\D t}}
\newcommand{\Dx}[1]{\frac {\D #1} {\D x}}
\newcommand{\Dy}[1]{\frac {\D #1} {\D y}}
\newcommand{\Dz}[1]{\frac {\D #1} {\D z}}

\newcommand{\mean}[1]{{\bar {#1}}}
\newcommand{\mmean}[1]{{\overline {#1}}}
\newcommand{\dev}[1]{{{#1}'}}
\newcommand{\inner}[2]{{\left<{#1,\ #2}\right>}}

\newcommand{\transfer}{{\Gamma}}
\newcommand{\Reynolds}{{\cal R}}  

\newcommand{\bracket}[1]{\left[#1\right]}
\newcommand{\cbrace}[1]{\left\{#1\right\}}
\newcommand{\parenth}[1]{\left(#1\right)}
\newcommand{\win}[2] {{{\bracket {#1}}_{#2}}}

\DeclareMathSymbol{\R}{\mathbin}{AMSb}{"52}

\begin{document}
\preprint{Preprint submitted to {\it Physics of Fluids}}

	\title{Perfect energy transfer and mean-eddy interaction 
		in incompressible fluid flows}

	\author{X. San Liang}
	\email{sanliang@cims.nyu.edu}
	\affiliation{Division of Engineering and Applied Sciences,
		Harvard University, Cambridge, Massachusetts}
	\affiliation{Courant Institute of Mathematical Sciences\\ 
		     New York, New York}

  	\date{October 16, 2006}

\begin{abstract}

The mean-eddy interaction in a flow is investigated in terms of the energy 
transfer between its mean and eddy processes. 
In the Reynolds decomposition framework, the classical energetics 
formalism does not have the transfer faithfully represented:
Energy transferred from the mean process to the eddy process
is not equal in amount to the energy transferred in the opposite direction.
As a result, the classical ``Reynolds stress extraction'',
$\Reynolds = - \mmean{\ve v' \ve v'} : \nabla \mean{\ve v}$,  
cannot be used to measure the mean-eddy interaction, or
turbulence production/laminarization in the context of turbulence research.
This paper shows that a faithful formalism 
can be rigorously established within the same framework.
The resulting transfer sums to zero everywhere over the 
decomposed subspaces, representing a mere redistribution 
of energy between the mean and eddy processes,
without generating or destroying energy as a whole.
For this reason, it has been referred to as {\it perfect transfer}
in distinction from other energy transfers one may have encountered.
The perfect transfer has a simple form in expression, with the introduction
of an eddy flow coupled with the mean and eddy processes of the 
field under concern.  The eddy or turbulence production is then totally
determined by the divergence of this flow.
This formalism has been validated with a well-known barotropic instability
model, the Kuo model for the stability of the zonal atmospheric jet stream.
We see a distribution of perfect transfer consistent with the instability
scenario inferred based on Kuo's theorem, while the traditional Reynolds
stress extraction $\Reynolds$ does not agree with the inference.
The formalism has also been validated with a wake control problem.
It is found that placement of a control to defeat the positive
perfect transfer $\transfer$ (mean-to-eddy transfer)
will yield the most efficient control
in terms of energy saving, in comparison to the inefficient control placement
based on $\Reynolds$ or eddy energy growth.
This research is expected to be useful for the harnessing of turbulence,
in that it may help to identify the best locations to place
passive controls, or design the performance functional for active controls.  

\end{abstract}



\maketitle

\section{Introduction}

\begin{figure} [h]
   \begin{center}
   \includegraphics[angle=0,width=0.5\textwidth]
   {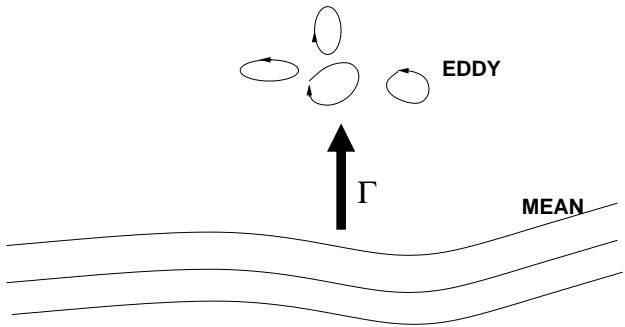}
   \caption{A schematic of the mean-eddy interaction,
	which is characterized by the energy transfer $\transfer$
	between the mean and eddy processes.
        \protect{\label{fig:schematic}}}
   \end{center}
   \end{figure}

The mean-eddy interaction in fluid flows is an important problem
in fluid mechanics. Related to it are 
hydrodynamic stability, turbulence production, laminarization, 
atmospheric cyclogenesis, hurricane generation, ocean eddy shedding, 
to name but a few. Central to the problem is the transfer of energy
between the mean and eddy processes as decomposed
(cf.~Fig.~\ref{fig:schematic}). 
The purpose of this paper is to quantify this transfer within 
the traditional Reynolds decomposition framework, and use it 
to investigate a new strategy of fluid control.
In a forthcoming paper, this formalism will be extended to
a more generic framework for real-time problems 
(Liang et al., manuscript submitted to SIAM J. Multiscale Model. Simul.)

The classical formalism of energy transfer can be best illustrated 
with the Reynolds decomposed equations for the advection of a scalar
field $T = \mean T + T'$ in an incompressible flow $\ve v$,
where the overbar stands for an ensemble mean, and the prime for the departure 
from the mean. In the absence of diffusion, $T$ evolves as
	\begin{eqnarray}
	\Dt T + \nabla\cdot(\ve v T) = 0,
	\end{eqnarray}
whose decomposed equations are
\begin{subequations}
\begin{eqnarray}
&&\Dt {\mean T} + \nabla\cdot(\mean{\ve v} \mean T 
		+ \mmean{\dev{\ve v} \dev T}) = 0, 	\label{eq:Tmean}	\\
 &&\Dt {\dev T} + \nabla\cdot(\dev {\ve v} \mean T + \mean {\ve v} \dev T
	        + \dev {\ve v}\dev T - \mmean{\dev {\ve v} \dev T}) = 0. 
							\label{eq:Teddy}	
\end{eqnarray}
\end{subequations}
Multiplying (\ref{eq:Tmean}) by $\mean T$, and (\ref{eq:Teddy}) by $\dev T$,
and taking the mean, one arrives at the evolutions of the mean energy
and eddy energy (variance)\cite{Lesieur}\cite{McComb}
	\begin{subequations}	\label{eq:TE}
	\begin{eqnarray}
	&& \Dt {{\mean T}^2/2} + \nabla\cdot  (\mean{\ve v} {\mean T}^2/2)
		= - \mean T \nabla \cdot (\mmean{\dev{\ve v} \dev T})	
			\label{eq:TEmean}	\\
	&& \Dt {\mmean{{\dev T}^2/2}} + \nabla\cdot  (\mmean{\ve v {\dev T}^2/2})
		= - \mmean {\dev{\ve v} \dev T}  \cdot \nabla \mean T.
			\label{eq:TEeddy}
	\end{eqnarray}
	\end{subequations}
The terms in divergence form are generally understood as the transports of
the mean and eddy energies, and those on the right hand side
as the respective energy transfers. The latter are usually used to explain 
the mean-eddy interaction.  Particularly, when $T$ is a velocity component,
the right hand side of (\ref{eq:TEeddy}) has been interpreted as
the rate of energy extracted by Reynolds stress, or ``Reynolds stress
extraction'' for short, against the mean field 
to fuel the eddy growth; in the context of turbulence research, 
it is also referred to as the ``rate of turbulence production''.
 
An observation of the two ``transfer terms'' on the right hand sides of
(\ref{eq:TE}) is that they are not symmetric; in other words, 
they do not cancel out each other.
In fact, they sum to $\nabla\cdot (\mean T \mmean{\dev{\ve v}\dev T})$,
which in general does not vanish.
This is not what one expects, as physically a transfer process
should be a mere redistribution of energy between the mean and eddy processes,
without destroying or generating energy as a whole.
These two quantities therefore are not real transfers, and
cannot be used to measure the mean-eddy interaction.

The reason for the asymmetry between the terms on the right hand side
of (\ref{eq:TE}) is that they are intertwined with transport processes;
or alternatively, the divergence terms on the left hand side do not account
for all the fluxes.
Some people such as Pope\cite{Pope} add an extra term in the flux term of 
(\ref{eq:TEmean}) to maintain the balance, but it is not clear how that term
should be chosen on physical grounds.
Pedlosky\cite{Pedlosky} pointed out that a partial solution of the problem
is to take averages for these terms over a substantially large domain.
This way the transport contributions may be reduced and hence
the transfer stands out.
Liang and Robinson\cite{LR1} argued that spatial averages should be avoided 
to retain the information of spatial intermittency in the energetics.
They believed that a precise separation between the transport and transfer
can be made to satisfy the above symmetric requirement. They even named
the transfer thus obtained {\it perfect transfer}, in distinction from other 
transfers that may have been called in the literature. 
But in their paper a rigorous formalization was postponed to future work;
how the separation can be achieved is still open,

This study intends to give this problem a solution in the traditional
framework. A complete answer to the issue of separation raised 
in \cite{LR1}, which is based on a new mathematical apparatus,
the {\it multiscale window transform} developed by Liang and Anderson
(manuscript submitted to SIAM J. Multiscale Model. Simul.),
is deferred to the sequel to this paper.
The following two sections are devoted to the establishment 
of a rigorous formalism for the transfer $\transfer$. 
We first consider the case for a scalar field $T$
(\ref{sect:scalar}), and then extend to momentum equations (\ref{sect:mom}). 
The formalism is validated with a well-studied 
instability model of an atmospheric jet stream (section~\ref{sect:kuo}),
and applied to harness the Karman vortex street behind a circular cylinder 
(\ref{sect:wake}). A brief summary is presented in section~\ref{sect:summary}.

\section{Formalism with a passive scalar}	\label{sect:scalar}

\subsection{Reynolds decomposition}	\label{sect:mathframework}

The transfer is sought within the Reynolds decomposition framework.
The key of the Reynolds decomposition is Reynolds averaging. It decomposes 
a field, for example a scalar field $T$, into a mean $\mean T$ plus a
departure from the mean, $T'$. 
Simple as it is, Reynolds averaging actually introduces an important geometric
structure which, as we will see shortly, helps to make the transfer problem
easier.

A Reynolds average may be understood either as an ensemble mean, 
or an expectation with respect to the measure of probability. 
Practically it may also be understood as an average in time 
or an average in some dimension of space. 
Its basic properties include: (1) $\mmean {T'} = 0$,
(2) $\mmean{\alpha T} = \alpha \mean T$, for $\alpha=\rm const$,
and a corollary from the above two, 
(3) $\mmean{\mean T T'} = \mean T \mmean{(T')} = 0$.
To put all these understandings together, 
the decomposition can be recast in the framework of a Hilbert
space $\cal H$, with an inner product defined as,
	\begin{eqnarray}
	\inner f g = \mmean {f g},
	\end{eqnarray}
for any $f$ and $g$ in the space $\cal H$, that is,
the ensemble, the probability space, or the space of functions
over the time or spatial domain under consideration,
if the overbar is, respectively, an ensemble mean, a probability expectation, 
or a time/spatial average. 
    (It is interesting to note that the meaning of a Reynolds averaging 
    is two-fold: one is the mean state reconstruction,
    another the summation or integration operator
    in forming the inner product.)

A Reynolds decomposition thus splits $\cal H$ into two subspaces,
which contain the mean process and the eddy process.
We will refer to these subspaces as {\it windows}, so we
have a mean window and an eddy window. 
Distinguish them respectively with subscripts $0$ and $1$. 
Correspondingly the decomposed components
of a field $T$, $\mean T$ and $T'$, will be alternatively 
written as $\win T 0$ and $\win T 1$ for convenience.
Using these notations, the energy of $T$ on window $k$ 
as defined in (\ref{eq:TE})
is $\frac 1 2 \inner {\win T k} {\win T k}$;
the property $\mmean{\mean T T'} = 0$ becomes 
$\inner {\win T 0} {\win T 1} = 0$, 
implying the two windows are orthogonal.
The concept of orthogonal windows puts 
the mean and eddy fields on the same footing, 
and will help to greatly simplify the derivation.

\subsection{Multiscale flux}

An important step toward the solution of the
transfer problem is finding the fluxes, 
and hence the transports, on the two scale windows.
The ($\mean{\ve v} \mean T^2/2 $) and ($\mmean{\ve v T'^2/2}$) 
in (\ref{eq:TEmean}) and (\ref{eq:TEeddy}), though seemingly
in flux forms, are not really the desiderata in a rigorous physical sense.
One may see this through a simple argument of energy conservation,
which requires that the
mean and eddy fluxes sum to $(\mmean{\ve v T^2/2})$--clearly 
these two quantities do not meet the requirement.

On the other hand, the concept of multiscale flux can be naturally 
introduced within the formalized Reynolds decomposition framework. 
Given a flow $\ve v$, the flux of an inner product 
$\inner f g$ over $\cal H$ is $\inner {\ve v f} g =  \inner f {\ve v g}$ 
($\ve v$ is self-adjoint with respect to $\inner\cdot\cdot$).
(Note the flux is uniquely represented this way. The only other choice one might
propose for the representation is $\inner {\ve v} {fg}$. 
This, however, does not make sense in physics,
as a flow $\ve v$ is an operator, not a function of the same like 
as $f$ and $g$ in the functional space.)
Let $f=g= \frac T {\sqrt 2}$, one obtains the flux
of energy 
	\begin{eqnarray}	\label{eq:flux}
	\ve Q = \frac 1 2 \inner{\ve v T} T.
	\end{eqnarray}
Geometrically, the right hand side of (\ref{eq:flux}) shows
a projection of $(\ve v T)$ onto $T$.
The flux on window~$k$ then should be a projection of $\ve v T$
onto $\win T k$:
	\begin{eqnarray}	\label{eq:Qk}
	\ve Q_k = \frac 1 2 \inner {\ve v T} {\win T k}
		= \frac 1 2 \inner {\win {\ve v T} k} {\win T k}.
	\end{eqnarray}
In arriving at the last result we have used the fact that the
two windows are orthogonal. 

The multiscale fluxes $\ve Q_k$ thus obtained are additive, i.e.,
$\ve Q = \ve Q_0 + \ve Q_1$. 
In fact, by the orthogonality between the mean and eddy windows, 
we immediately have
	$$\sum_{k=0}^1 \frac 1 2 \inner{\win {\ve v T} k} {\win T k} 
 		= \frac 1 2 \inner{\ve v T} {T}.$$ 
This is the very conservation requirement mentioned above.

\subsection{Perfect transfer}	\label{sect:perfect}

Continue to examine the evolution of $T$ in an incompressible flow $\ve v$.
In the language introduced in subsection~\ref{sect:mathframework},
Eqs.~(\ref{eq:Tmean}) and (\ref{eq:Teddy}) can 
be written in a unified form:
	\begin{eqnarray}	\label{eq:Tk}
	\Dt {\win T k} + \nabla\cdot \win {\ve v T} k = 0,
	\end{eqnarray}
for windows $k=0, 1$.
Here the decomposition is performed only in a statistical sense.
That is to say, the mean is an ensemble mean or a probability 
expectation. But as we will see toward the end of this section,
the formalism of energy transfer is essentially the same with
respect to other methods of averaging.

Application of $\inner {\win T k} \cdot $ to (\ref{eq:Tk}) gives
the energy evolution equation on window~$k$:
	\begin{eqnarray}	\label{eq:TEk}
	\Dt {\frac 1 2 \inner {\win T k} {\win T k}} + 
	\inner {\win T k} {\nabla\cdot \win {\ve v T} k} = 0.
	\end{eqnarray}
%
%
The nonlinear term (second part on the l.h.s) 
involves two interwoven processes: transport and transfer.
The former integrates to zero over a closed spatial domain; 
the latter sums to zero over $k$, $k=\{0,1\}$. That is to say,
(\ref{eq:TEk}) can be symbolically written as,
	\begin{eqnarray}	\label{eq:TEk2}
	\Dt {\frac 1 2 \inner{\win T k} {\win T k} }
	= - \nabla\cdot \ve Q_k + \transfer_k,
	\end{eqnarray}
where $\ve Q_k$ is the flux on window $k$, and $\transfer_k$ the transfer to window
$k$ from its complementary subspace.

We already know the multiscale flux $\ve Q_k$ in (\ref{eq:Qk}).
The transfer $\transfer_k$ is now easy to derive. 
Comparing (\ref{eq:TEk2}) with (\ref{eq:TEk}), one obtains
	\begin{eqnarray}
	\transfer_k - \nabla\cdot\ve Q_k = 
		- \inner {\win T k} {\nabla\cdot \win{\ve v T} k}.
	\end{eqnarray} 
Substitution of (\ref{eq:Qk}) immediately gives
	\begin{eqnarray}	\label{eq:trans1}
	\transfer_k = \frac 1 2 \nabla\cdot \inner {\win {\ve v T} k} {\win T k}
			- \inner {\win T k} {\nabla\cdot\win {\ve v T} k}.
	\end{eqnarray} 

It would be more clear to see the mean-eddy interaction if
(\ref{eq:trans1}) is rewritten in the traditional overbar/prime notation.
For the eddy process ($k=1$), the transfer from the mean flow is
	\begin{eqnarray}
	\transfer_1 
	= \frac 1 2 \nabla\cdot \parenth{\mmean{\dev {(\ve v T)} \dev T}}
		    - \mmean{\dev T \nabla\cdot \dev {(\ve v T)}},
	\end{eqnarray}
which reduces to 
    \begin{center}
      \framebox[0.50\textwidth]{
      \begin{minipage}[c]{1\textwidth}
        \begin{eqnarray}
	\transfer_1 \equiv \transfer
		  = \frac 1 2 \cbrace{	
		    \mean T \nabla\cdot (\mmean{\dev{\ve v} \dev T})
		  - (\mmean {\dev{\ve v} \dev T}) \cdot \nabla \mean T
		      }.		 \label{eq:transfer}\\
						 \nonumber
        \end{eqnarray}
      \end{minipage}
      }
    \end{center}
In the derivation, the incompressibility assumption $\nabla\cdot\ve v=0$,
and hence $\nabla\cdot\mean{\ve v} = 0$, $\nabla\cdot\ve v' = 0$,
has been used. Likewise, 
	\begin{eqnarray}	\label{eq:transfer0}
	\transfer_0 = \frac 1 2 \nabla\cdot \mmean{(\mmean{(\ve v T)} \mean T)}
			- \mmean {\mean T \nabla\cdot {(\ve v T)} }
		    = - \transfer,
	\end{eqnarray}
and
	\begin{eqnarray}
	\ve Q_0 &=& \frac 1 2 \cbrace{\mean {\ve v} {\mean T}^2 
			    + \mean T \mmean{\ve v'T'}},  \label{eq:Q0} \\
	\ve Q_1 &=& \frac 1 2 \cbrace{\mmean{\ve v T'^2} 
				+ \mean T \mmean{\ve v'T'} }
	\end{eqnarray}
The mean-eddy energetics corresponding to (\ref{eq:TE}) are, therefore, 
	\begin{subequations}	\label{eq:TE2}
	\begin{eqnarray}
	&& \Dt {{\mean T}^2/2} + \nabla\cdot \parenth{
		\frac 1 2 \mean {\ve v} {\mean T}^2 + 
		\frac 1 2 \mean T \mmean{\ve v'T'}}
		= - \transfer, \label{eq:TE2mean}	\\
	&& 
	   \Dt {\mmean{{\dev T}^2/2}} + \nabla\cdot \parenth{
		\frac 1 2 \mmean{\ve v T'^2} +
		\frac 1 2 \mean T \mmean{\ve v'T'}}
		 = \transfer,
					\label{eq:TE2eddy}
	\end{eqnarray}
	\end{subequations}
with $\transfer$ as shown in (\ref{eq:transfer}).

Equations (\ref{eq:transfer}) and (\ref{eq:transfer0}) imply an important
property for the transfer derived above,
    \begin{center}
      \framebox[0.25\textwidth]{
      \begin{minipage}[c]{1\textwidth}
	\begin{eqnarray}
		\sum_k \transfer_k = 0.	\label{eq:perfect}\\
					\nonumber
	\end{eqnarray}
      \end{minipage}
      }
    \end{center}
That is to say, the transfer thus obtained is a process of energy 
redistribution between the windows; there is no energy generated 
or destroyed as a whole, just as one may expect.
To distinguish from other energy transfers one may have encountered 
in the literature, we will refer to $\transfer$ as {\it perfect
transfer}, a term adopted from \cite{LR1}, when confusion might arise.

Note the distinct difference between $\transfer$ and the Reynolds
stress extraction as appears in (\ref{eq:TEeddy}), 
	$\Reynolds = - \mmean {\dev{\ve v} \dev T}  \cdot \nabla \mean T$,
which traditionally has been used to interpret the generation of eddy events,
and has been interpreted as, in the turbulence research context, 
the turbulence production rate. 
In sections~\ref{sect:kuo} and \ref{sect:wake}, 
we will see that these two are in general differently
distributed in space and time.

The transfer $\transfer$ may be further simplified in expression. 
If $\mean T \ne 0$, (\ref{eq:transfer}) may be alternatively written as
	\begin{eqnarray}	\label{eq:trans2}
	\transfer = \frac 1 2 {\mean T}^2\ 
		\nabla\cdot \parenth{\frac {\mmean{\ve v'T'}} {\mean T}}.
	\end{eqnarray}
Observe that the quantity in the parenthesis has the dimension of velocity.
It represents a flow coupled with the mean and eddy processes of $T$.
For convenience, introduce a ``$T$-coupled eddy velocity''
	\begin{eqnarray}	\label{eq:Tvelo}
	\ve v_T = {\frac {\mmean{\ve v'T'}} {\mean T}},
	\end{eqnarray}
then  
	\begin{eqnarray}	\label{eq:transfer2}
	\transfer = \frac 1 2 \mean T^2 \nabla\cdot \ve v_T.
	\end{eqnarray}
Notice that $\frac 1 2 \mean T^2$ is the mean energy of $T$ and is hence 
always positive, so whether eddies are produced is totally determined by the 
divergence of the $T$-coupled eddy flow.

The $T$-coupled eddy flow $\ve v_T$ is introduced for notational simplicity 
and for physical understanding. One should be aware that $\transfer$ is 
well defined, even though $\ve v_T$ does not exist when $\mean T=0$.
In that case, the original expression (\ref{eq:transfer}) should be used.

\subsection{Other methods of averaging}

Practically the Reynolds averaging is often performed with respect
to time if the process is stationary, or some dimension of space
if the process in homogeneous in that dimension.
If the averaging is in time, the above derivations also apply,
except that the time derivatives in 
(\ref{eq:Tk}), (\ref{eq:TEk}), (\ref{eq:TE2mean}), and
(\ref{eq:TE2eddy}) are gone.
The transfer $\transfer$ is still in the same form as (\ref{eq:transfer}).

If the averaging is performed in a dimension of space, say, in $x$,
then the above derivation needs modification, as the averaging 
does not commute with $\frac \D {\D x}$. But we have the following 
extra properties:
   $\mmean {\frac {\D \psi}{\D x}} = 0$, $\frac {\D \mmean {\psi }} {\D x} = 0$,
for any field $\psi$. These substituted into the continuity equation 
$\nabla\cdot\ve v = 0$ yield 
$\Dx {\mean u} = \Dy {\mean v} + \Dz {\mean w} = 0$.
With these identities, we repeat the procedures in the above subsection,
and obtain:
	\begin{eqnarray}	\label{eq:tmp1}
	\transfer = \frac 1 2 \cbrace{	
	    \mean T \nabla_{yz} \cdot (\mmean{\dev{\ve v} \dev T})
	  - (\mmean {\dev{\ve v} \dev T}) \cdot \nabla_{yz} \mean T}.
	\end{eqnarray}
Here $\nabla_{yz} = \ve j \Dy {\ } + \ve k \Dz {\ }$ is the 
$\nabla$ operator with the $x$ component removed.

Notice that $\mmean{u' T'}$ and $\mean T$ are independent of $x$,
viz
	\begin{eqnarray}	\label{eq:tmp2}
	\frac 1 2 \parenth{	
		\mean T \Dx {\mmean{u'T'}} -
		\mmean{(u'T')} \Dx {\mean T}}
		= 0.
	\end{eqnarray}
So we may add the left hand side of (\ref{eq:tmp2}) to (\ref{eq:tmp1})
to get
	\begin{eqnarray}	
	\transfer = \frac 1 2 \cbrace{	
	    \mean T \nabla \cdot (\mmean{\dev{\ve v} \dev T})
	  - (\mmean {\dev{\ve v} \dev T}) \cdot \nabla \mean T},
	\end{eqnarray}
which is precisely the same as (\ref{eq:transfer}) in expression form.

In a brief summary, we have derived the mean-eddy energy transfer 
for a passive scalar in an incompressible flow, which is ``perfect''
in the sense that the energy extracted from the mean is equal in
amount to the energy released by the mean.
The perfect transfer is invariant in expression form 
with averaging schemes.

\section{Formalism with momentum equations}	 \label{sect:mom}

We need to deal with momentum equations for the mean-eddy kinetic energy 
transfer. Consider an incompressible ideal flow $\ve v$. 
It is governed by
	\begin{subequations}		\label{eq:gov}
	\begin{eqnarray}	
	&& \Dt{\ve v} + \nabla\cdot (\ve v \ve v) = - \nabla P,	\label{eq:mom} \\
	&& \nabla \cdot \ve v = 0,				\label{eq:cont}
	\end{eqnarray}
	\end{subequations}
In forming the multiscale energy equations, the pressure term only
contributes to the transport, i.e., the resulting pressure work is
in a divergence form, thanks to the incompressibility assumption.
So only the nonlinear terms require some thought in deriving
the perfect transfer.

In this case, the formalism with the momentum equations is then 
essentially the same as that with the evolution of a passive scalar,
except that we now have three ``scalars'' for the three velocity
components $u$, $v$, and $w$.
If $\mean u \ne 0$, $\mean v \ne 0$, and $\mean w\ne 0$,
for each component, there is a ``coupled eddy velocity'' defined
by (\ref{eq:Tvelo}), so we have $\ve v_u$, $\ve v_v$, and $\ve v_w$,
which are         
		${\frac {\mmean{\ve v'u'}} {\mean u}}$,
		${\frac {\mmean{\ve v'v'}} {\mean v}}$, and
		${\frac {\mmean{\ve v'w'}} {\mean w}}$,
respectively.
According to the preceding section, each velocity corresponds to 
a transfer as expressed by (\ref{eq:transfer2}).
The total kinetic energy transfer is thence the sum of all 
the three transfers:
	\begin{eqnarray}
	\transfer = \frac 1 2 \cbrace{
		\mean u^2 \nabla\cdot {\ve v}_u +
		\mean v^2 \nabla\cdot {\ve v}_v +
		\mean w^2 \nabla\cdot {\ve v}_w }.
	\end{eqnarray}

As noted in subsection~\ref{sect:perfect}, the above formula is well 
defined even when the mean velocity vanishes. In that case, one just 
needs to expand it to obtain:
	\begin{eqnarray}	\label{eq:transfervec}
	\transfer = \frac 1 2 \cbrace{
		\nabla\cdot (\mmean{\ve v' \ve v'}) \cdot \mean{\ve v}
		- (\mmean{\ve v' \ve v'}) : \nabla\mean{\ve v} },
	\end{eqnarray}
where the second term in the curly braces is the very 
Reynolds stress extraction:	
	\begin{eqnarray}	\label{eq:Reynoldsvec}
	\Reynolds = - (\mmean{\ve v' \ve v'}) : \nabla\mean{\ve v}.
	\end{eqnarray}
Like those with a scalar field, these formulas stay invariant in form,
no matter what an averaging scheme is adopted.

\section{Validation with an instability model} \label{sect:kuo}

In this section, the above formalism of perfect transfer $\transfer$
is validated with an idealized instability model. We will
also see through this concrete example how $\transfer$ differs 
from the classical Reynolds stress extraction against 
the basic profile.

Consider a well-studied barotropic instability model, the
Kuo model, for the instability of the zonal atmospheric jet 
stream\cite{Kuo}\cite{Kuo2}. Liang and Robinson\cite{LR2} have constructed
a particular solution with a highly localized structure which
is ideal for our purpose here.
In the following we briefly present this solution, and
then calculate the transfer (\ref{eq:transfervec}).

Choose a coordinate frame 
with $x$ pointing eastward, $y$ northward. 
The governing equations for the Kuo model are the 2D version of (\ref{eq:gov}),
but with a Coriolis force term $f\ve k \wedge \ve v$ ($f$ constant) 
on the left hand side. The domain is periodic in $x$, and limited
within latitudes $y = \pm L$, where a slip boundary condition 
$v = 0$ is applied.
As rotation makes no contribution to the energy evolution,
the formulas established in section~(\ref{sect:mom}) equally apply here,
i.e., the Kuo model can be used for the validation.

\begin{figure} [h]
   \begin{center}
   \includegraphics[angle=0,width=0.75\textwidth]
   {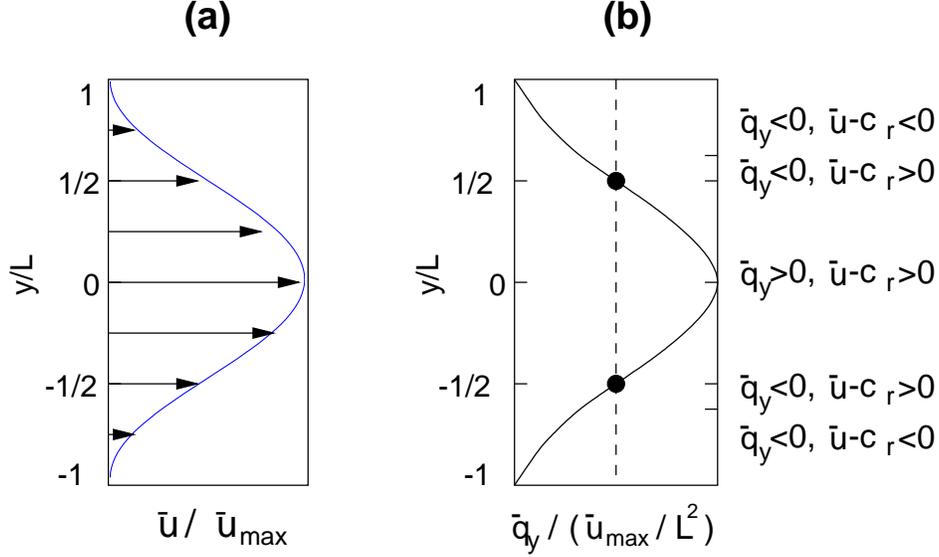}
   \caption{Configuration of the Kuo model.
           (a) The basic flow profile $\bar u = \bar u(y)$.
           (b) The background potential vorticity. Marked
                are the two reflection points on the profile curve.
        \protect{\label{fig:Kuo1}}}
   \end{center}
   \end{figure}

Assume a basic velocity profile 
(cf.~Fig.~\ref{fig:Kuo1}a)
        \begin{eqnarray}
        \bar u(y) = {\bar u}_{\max} \cos^2 \parenth{\frac\pi 2 \frac y L},
	\qquad {\mean u}_{\max} > 0.
        \end{eqnarray}
The background potential vorticity $q$ has
a meridional gradient (cf.~Fig.~\ref{fig:Kuo1}b)
        \begin{eqnarray}
        {\bar q}_y = - {\bar u}_{yy}
                   = - \frac {\pi^2} {2L^2} {\bar u}_{\max} \cos\frac {\pi y} L,
        \end{eqnarray}
which changes sign at $y = \pm \frac L 2$, meeting the necessary 
condition for instability by Rayleigh's theorem ({\it ibid}). 
Decompose the flow as 
	\begin{eqnarray}
	(u,v) = (\mean u(y), 0) + (u', v'),
	\end{eqnarray}
and substitute back to the governing equations.
Kuo considered only the initial stage of instability when
the perturbation field $(u',v')$ is very small. So the resulting
equations can be linearized. Assuming a solution of
the form
	\begin{eqnarray}	\label{eq:kuosolform}
	(u',v') = (\tilde u(y), \tilde v(y)) e^{ik(x - ct)},
	\end{eqnarray}
one obtains an eigenvalue problem 
	\begin{eqnarray}	\label{eq:eigen}
	\frac {d^2 \tilde v} {d y^2} + 
	\parenth{\frac {\mean u_yy} {c - \mean u} - k^2} \tilde v = 0,
	\end{eqnarray}
with boundary conditions
	\begin{eqnarray}
	\tilde v = 0, \qquad\qquad {\rm at}\ y=\pm L.
	\end{eqnarray}
The solution of (\ref{eq:eigen}) is not repeated here;
the reader may refer to Kuo's original papers for details.

Kuo showed that, in addition to the $q_y$ inflection requirement,
the difference $(\mean u - c_r)$ 
($c_r = {\rm Re} \{c\}$ the mode phase velocity) must be 
positively correlated with $\mean q_y$ over $[-L,L]$
in order for the perturbation to destabilize the jet.
In other words, for an instability to occur, it requires that
	\begin{itemize}
	\item[(1)] $\mean q_y$ change sign through $y\in[-L,L]$ (Rayleigh' theorem);
	\item[(2)] $(\mean u - c_r)$ and $\mean q_y$ be positively correlated 
		over $[-L,L]$ (Kuo's theorem).
	\end{itemize}
Hence the zero points of $(\mean u - c_r)$ and $\mean q_y$ are critical.
We will validate our transfer formalism through examining the instability
structures near these critical points. 
We choose a particular unstable mode (and hence a particular $c_r$)
to fulfill the objective.

As shown in \cite{LR2}, the wavenumber
$k = \frac 3 4 \frac \pi L$ gives such a mode;
it lies within the unstable regime as computed by Kuo\cite{Kuo2}.
In fact, if substituting back into the eigenvalue problem,
one obtains, using the shooting method\cite{Verterling},
	\begin{eqnarray}
	c = c_r + i c_i = (0.4504 + 0.0476i) {\mean u}_{\max},
	\end{eqnarray}
yielding a positive growth rate $kc_i>0$. 
Solved in the mean time is the corresponding eigenvector $\tilde v$,
which substituted in (\ref{eq:kuosolform}) and the governing equations
give a solution of all the fields.
The resulting phase speed $c_r = 0.4504 \mean u_{\max}$ 
and the gradient of the basic potential vorticity $\mean q_y$
give four critical values of $y$:
	\begin{eqnarray}
	\begin{array} {ll}
	\mean u(y) - c_r = 0		&\Longrightarrow y=\pm 0.53L,\\
	\mean q_y = - \mean u_{yy} = 0  &\Longrightarrow y=\pm 0.50L.
	\end{array}
	\end{eqnarray}

The four critical latitudes, as marked in Fig.~\ref{fig:Kuo1}b, 
partitions the $y$ dimension into five distinct regimes
characterized by different values of $K \equiv \mean q_y (\mean u - c_r)$.
For most of $y\in[-L,l]$, $K>0$, but the positivity 
is interrupted by two narrow strips 
near $y = \pm L/2$, where $K<0$.
This scenario has profound implications by Kuo's theorem.
Although Kuo's theorem is stated in a global form,
it should hold locally within the correlation scale.
In the present example, that means one of the necessary conditions
for barotropic instability is not met around the strips
and so there should be no instability occurring there.

   \begin{figure} [h]
   \begin{center}
   \includegraphics[angle=0,width=0.5\textwidth]
   {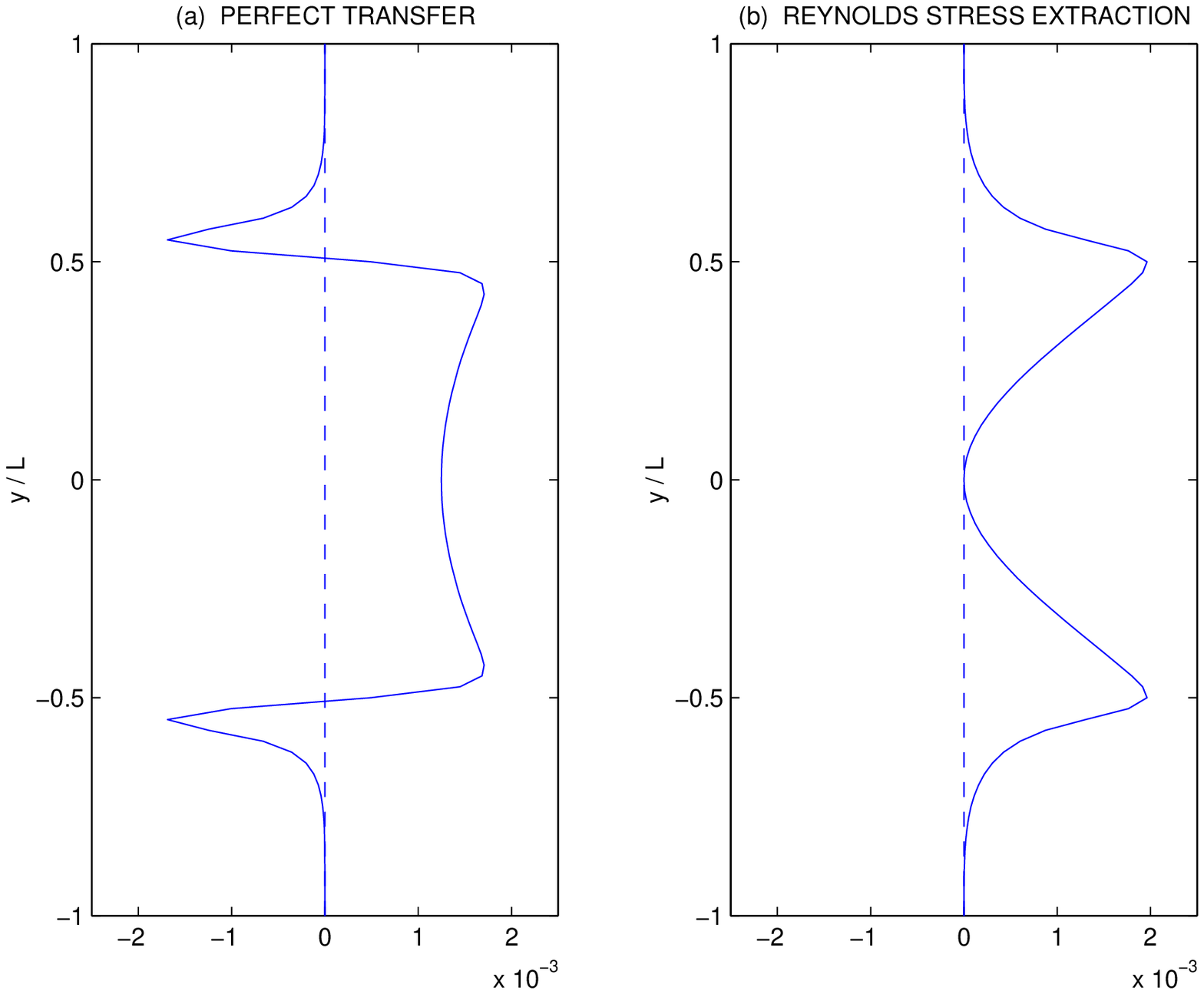}
   \caption{The barotropic energy transfer (scaled by ${\mean u}_{\max}^3 / L$)
	for the Kuo's model:	
	(a) the perfect transfer $\transfer$,
	    which is $\frac 1 2\bracket
		{\mean u \Dy {\mmean{u'v'}} - \mmean{u'v'} \Dy {\mean u}}$
	    here by (\ref{eq:transfervec}); 
	(b) the Reynolds stress extraction $\Reynolds$,
	    which is equal to $-\mmean{u'v'} \Dy {\mean u}$ here.
	The averaging is taken with respect to $x$.
	\protect{\label{fig:Kuo2}}}
   \end{center}
   \end{figure}

Instability means a transfer of energy from the background
to the perturbation field, namely, a positive $\transfer$.
Using the particular solution obtained above, we compute
the transfer from (\ref{eq:transfervec}). 
We adopt a zonal averaging, i.e., averaging in $x$, to fulfill
the decomposition. This is because, (1) $\mean u$ itself does not 
have $x$-dependence and hence can be understood as an $x$-average, 
and (2) the solution is homogeneous in $x$ due to the cyclic
boundary condition. The computation is straightforward.
The result is plotted in Fig.~\ref{fig:Kuo2}a.
Sure enough, $\transfer$ is not positive around the two narrow strips; 
in fact, there is a strong negative transfer, i.e., upscale or inverse 
transfer from the eddy window to the background.
Moreover, the negative transfer is limited within two narrow regimes,
just as one may expect by Kuo's theorem.
In contrast, a different scenario is seen on the profile of the 
conventional Reynolds stress extraction $\Reynolds$, 
which we plot in Fig.~\ref{fig:Kuo2}b. $\Reynolds$ is nonnegative
throughout $[-L,L]$; particularly, it is maximally positive
over the narrow strip regimes, countering our foregoing intuitive argument.
Through this example, our perfect transfer $\transfer$ results in a scenario
agreeing well with the analytical result of the Kuo model, 
while the conventional Reynolds stress extraction $\Reynolds$ does not.

\section{Application to the suppression of eddy shedding} \label{sect:wake}

A practical application of the above research is turbulence control.
Turbulence control is a technique to manipulate turbulence growth
to achieve the goal of drag reduction (cf.~\cite{Farrell}
and particularly the celebrated paper by Kim\cite{Kim}, and the references therein).
How the current research may come to help is to provide a better object, 
i.e., the perfect transfer, to manipulate, 
in place of the growth of turbulence energy or eddy energy.

\begin{figure}
   \begin{center}
   \includegraphics[angle=0,width=0.6\textwidth]
   {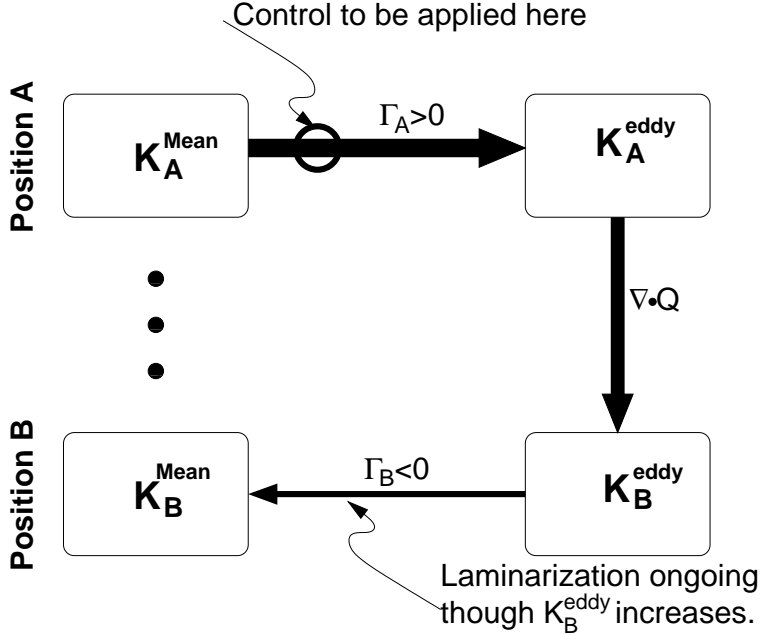}
   \caption
        {
         Schematic of the eddy energy transport and transfer for a
         two-point turbulent system. 
	 $K$ stands for kinetic energy, and
	 $\transfer$ and $\nabla\cdot\ve Q$ 
	 for transfer and transport, respectively. An arrow indicates the
         direction of an energy flow, with its thickness standing
         for strength. In this case, transfer is toward the
         mean at position B, but $K_B^{eddy}$ still grows
         because of the transport (advection) of $K_A^{eddy}$
         from position A. For optimal results, control
         should be placed at A only.
        \protect{\label{fig:ctr_schem}}}
   \end{center}
   \end{figure}

The proposal is out of the concern of how to maximally take advantage of
the processes of self-laminarization or relaminarization that may 
occur within a turbulent flow. In the interest of energy saving, 
suppression of the positive transfer $\transfer$ [cf.~(\ref{eq:transfervec})] 
is preferred to suppression of the eddy energy growth
to inhibit the production of turbulence.
To see why, observe that 
{\it eddy energy increase does not necessarily occur in accordance 
with a positive transfer}, and hence {\it a place where turbulence grows 
does not necessarily correspond to turbulence production}.
Actually, the correspondence is an exception rather than a rule.
(Later in this section we will see an example.)
The energy needed to fuel the growth could be transported from 
the neighborhood, rather than released {\it in situ}.
One possibility is, while disturbances rapidly grow,
a process of laminarization could be undergoing at 
the very position. As shown in the two-point system in Fig.~\ref{fig:ctr_schem},
while disturbances grow at both $A$ and $B$ (both $K_A^{eddy}$ and $K_B^{eddy}$
increase), the eddy energy is produced at A only. 
At B, not only there is no eddy energy production, 
but the transfer is from the eddy window
to the mean window. That is to say, the system is undergoing a laminarization 
at $B$, even though the eddy energy $K_B^{eddy}$ grows, because of
a surplus of the influx of eddy energy over the inverse
transfer. 
Control of the eddy energy growth at both A and B
indeed helps to suppress the onset of turbulence,
but it is not optimal in terms of energy saving.
Suppression of $K_B^{eddy}$ defeats the intrinsic trend of 
laminarization in the mean time,
and therefore reduces the control performance. To take advantage of this
laminarization, the control should be applied at position A,  i.e., 
the source region, only, and the optimal objective functional should be 
designed with respect to $\transfer_A$, 
rather than $K_A^{eddy}+K_B^{eddy}$.

In this spirit, we demonstrate the application by showing how 
one may efficiently suppress the vortex shedding behind a circular
cylinder. Vortex suppression is important in that it can result
in significant drag reduction and hence energy saving; it may also
be used to reduce noise.
Presented in the following are just some diagnostic results with 
a saturated wake to which the afore-established formalism is applicable; 
the same example will be studied in more detail 
in the sequel to this paper
with nonstationarity considered.
We will deal with a laminar case only, 
but the idea equally applies to turbulent wakes.

There are many sophisticated techniques to suppress the shedding of 
vortices in a wake [cf.~\cite{Huerre} and \cite{Oertel} 
and the references therein]. Surface-based suction is one of them.
To our knowledge, the research along this line thus far, however, has been 
focused on the technique per se. No report has been found on the issue
of where to place the suction to optimize the performance.
In the following, we will show that our formalism of mean-eddy 
interaction and perfect transfer can give this question an answer.

\begin{figure}
   \begin{center}
   \includegraphics[angle=0,width=0.6\textwidth]
   {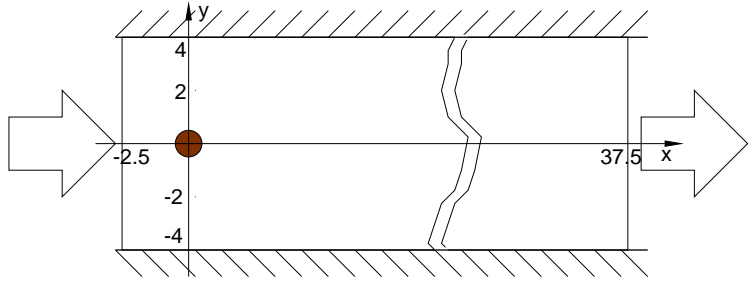}
  \caption
        {Model configuration. The coordinates
	$x$ and $y$ are scaled by the cylinder diameter $d$.
        \protect{\label{fig:ctr_domain}}}
   \end{center}
   \end{figure}

Consider a planar flow passing around a circular cylinder.
The governing equations are the same as those of Eqs.~(\ref{eq:gov}),
except that dissipation is included. The computational domain is
plotted in Fig.~\ref{fig:ctr_domain}, with $x$ and $y$ nondimensionalized
by the cylinder diameter $d$. A uniform inflow $(U,0)$ is specified at
$x=-2.5$, and on the right open boundary ($x=37.5$) a 
radiative condition\cite{Orlanski} is applied.
At $y=\pm 4$ are two solid boundaries, where nonslip conditions are
imposed. 

We examine a flow with Reynolds number $ Re = \frac {Ud} {\nu} = 200$.
The spacing choice of $\Delta x$ and $\Delta y$ is found
not a stringent constraint.
By experiments a mesh with $\Delta x = \Delta y = 0.05$
and a mesh with $\Delta x = \Delta y = 0.025$ produce little
difference in the final result for our problem.
We thus choose $\Delta x = \Delta y = 0.05$ for economic reason.
The governing equations are integrated forward on a staggered grid\cite{Arakawa}
using a semi-implicit (implicit for pressure) 
finite difference scheme (e.g., \cite{Kreiss}) 
of the second order in both time and space.

\begin{figure} [h]
   \begin{center}
   \includegraphics[angle=0,width=0.5\textwidth]
   {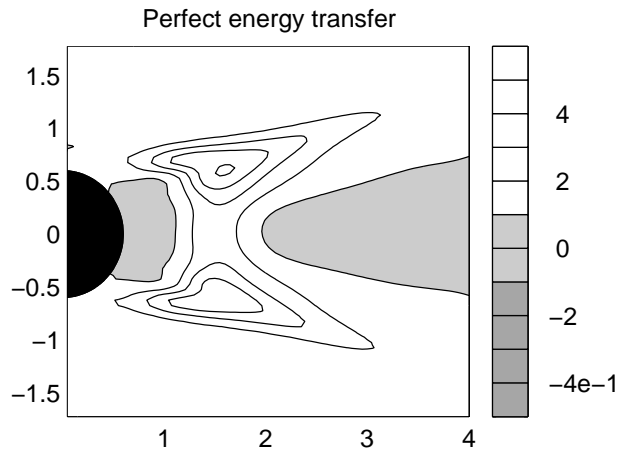}
   \caption{Perfect transfer $\transfer$ in the wake behind a cylinder
	(units in $U^3/d$). 
	\protect{\label{fig:ctr_transfer}}}
   \end{center}
   \end{figure}

The model is run until a statistical equilibrium is reached.
After that, it is integrated further for 100 time units and 
the outputs are used to calculate the transfer $\transfer$.
The stationarity in time makes it a natural choice to perform
time averaging in computing the transfer (\ref{eq:transfervec}).
The computation is straightforward.
We plot the result in Fig.~\ref{fig:ctr_transfer}.
Note the time interval is large enough that
one virtually sees no difference in the computed result
if it is enlarged.

\begin{figure} [h]
   \begin{center}
   \includegraphics[angle=0,width=1\textwidth]
   {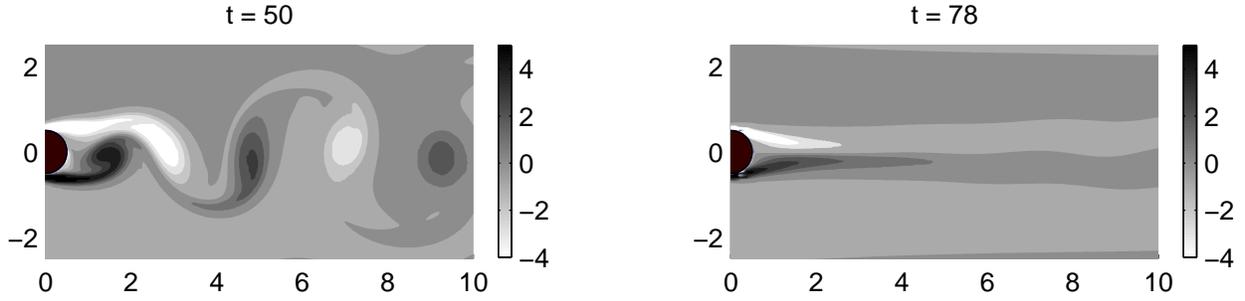}
   \caption{
	Snapshots of the vorticity in the optimal control experiment.
	The control is applied at $t=50$ and so forth.
	\protect{\label{fig:ctr_vort_suct}}}
   \end{center}
   \end{figure}

By (\ref{eq:transfervec}), a positive $\transfer$ means
eddy energy generation or turbulence production in turbulence research,
while a negative $\transfer$ indicates a transfer in the opposite direction.
In Fig.~\ref{fig:ctr_transfer}, two triangular lobes of strong
positive $\transfer$ sit on either side of the axis $y=0$, with a 
weak negative center lying in the near wake. That is to say, eddy energy 
is generated within the two lobes, while in between is a 
laminarization process. By the forgoing arguments, 
an efficient control strategy should be the one inhibiting the positive
$\transfer$ in these two lobes.
Since we are considering only the technique of surface-based suction,
the $\transfer$ distribution suggests that application of suctions near 
the two lobes should be effective. In doing this, one can simultaneously
take advantage of the laminarization process occurring in the near wake. 
Indeed, our control experiments show 
that the areas between 50 to 80 degrees and between $-50$ to $-80$ 
degrees from the x-axis are the effective suction locations 
to suppress the vortex street.
The effectiveness, according to \cite{Oertel}, may be measured by a 
suction rate
        $c_q = \frac m {U d}$, 		
where $m$ is the mass flow rate, $U$ the free stream velocity, 
and $d$ the cylinder diameter.
By experiments, the most effective control is that placed at 
$\pm 70^o$, following the same orientation of the two positive $\transfer$ 
lobes. Only a rate of $c_q = 0.18$ can have the vortices completely 
suppressed (e.g., Fig.~\ref{fig:ctr_vort_suct}).
In contrast, controls in areas below
50 degrees and above $-50$ degrees are counterproductive, 
as the near wake laminarization process is defeated.

\begin{figure} [h]
   \begin{center}
   \includegraphics[angle=0,width=1\textwidth]
   {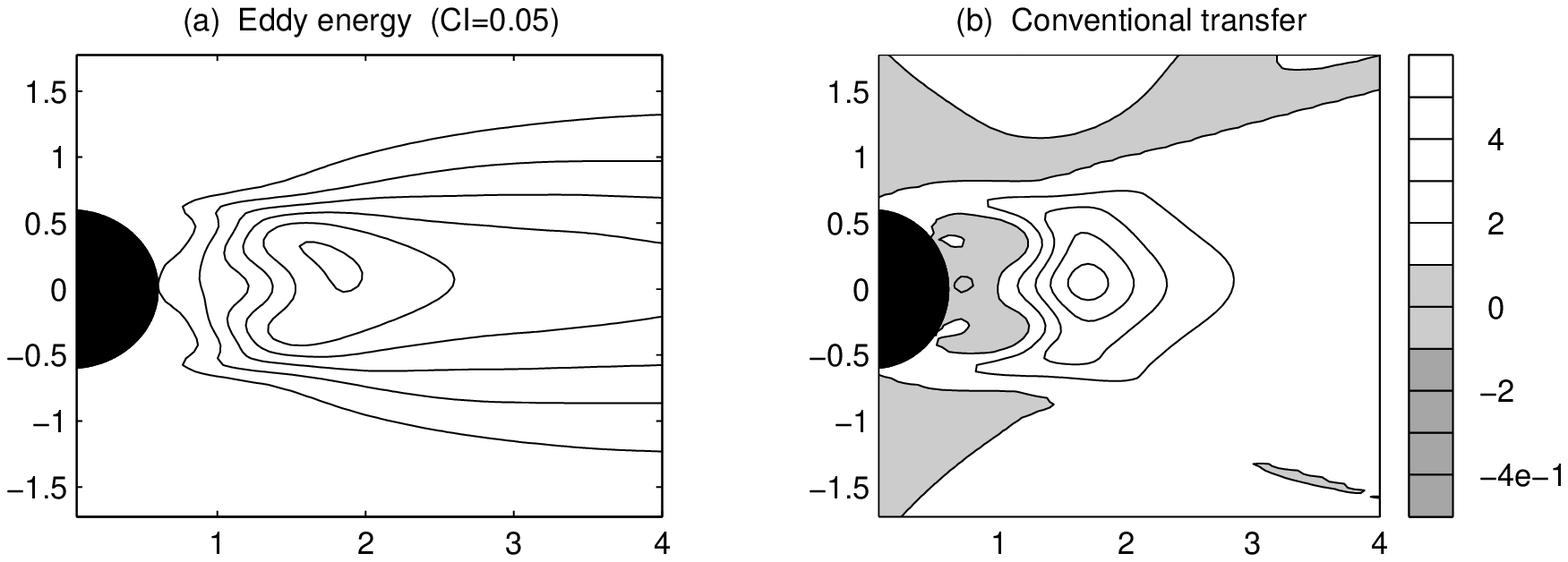}
   \caption{(a) Perturbation energy (in $U^2$); 
	    (b) $\Reynolds$ (in $U^3/d$).
	\protect{\label{fig:ctr_transfer2}}}
   \end{center}
   \end{figure}

It is of interest to see how other diagnostic fields, such as the eddy energy 
$\frac 1 2 \mmean{(u'^2 + v'^2)}$ and $\Reynolds$, are distributed.
Shown in Figs.~\ref{fig:ctr_transfer2}a and b are 
these fields. Clearly, they both attain their maxima along the axis $y=0$, 
a scenario completely different from that of $\transfer$
in Fig.~\ref{fig:ctr_transfer}. If the control is based on these fields,
a suction should be placed at $(x,y) = (\frac 1 2, 0)$, i.e.,
$0^{\rm o}$ from the axis. But, as noted above, the control experiments
show this does not result in an effective vortex suppression.

The success of the control experiment serves to validate our formalism 
of the mean-eddy interaction and the equations of $\transfer$, 
(\ref{eq:transfer}) and (\ref{eq:transfervec}).
In the mean time, a variety of fluid control problems, both
active and passive, may benefit from this formalism.

\section{Conclusions and discussion}  \label{sect:summary}

In the Reynolds decomposition framework, the mean-eddy interaction 
has been rigorously formulated in terms of energy transfer,
which can be singled out from the intertwined nonlinear processes 
by eliminating the transport effect.
The resulting transfer sums to zero over the two decomposed subapces,
or {\it windows} as called in the text.
In other words, the transfer represents a redistribution process 
between the mean and eddy windows, without generating or destroying
energy as a whole. Because of this property, it is sometimes
referred to as {\it perfect transfer} in distinction from other 
transfers that one may have encountered in the literature.

The perfect transfer from the mean process to the eddy process 
can be explicitly written out. 
In the case of a scalar $T$ advected by an incompressible flow $\ve v$,  
traditionally there is a quantity
	\begin{eqnarray*}
	\Reynolds = - (\mmean{\ve v'T'}) \cdot \nabla \mean T,
	\end{eqnarray*}
which, when $T$ is a velocity component, 
has been explained as the rate of energy extracted by 
Reynolds stress against the basic profile.
We showed that this is not the eddy energy transferred from 
the mean to the eddy windows. The real transfer should be
	\begin{eqnarray*}
	\transfer = \frac 1 2 \bracket{
		\mean T \nabla\cdot (\mmean{\ve v' T'}) 
		+ \Reynolds},
	\end{eqnarray*}
which may also be written as
	\begin{eqnarray*}
	\transfer = \frac 1 2 \mean T^2\ \nabla\cdot \ve v_T,
	\end{eqnarray*}
if $\mean T \ne 0$, in terms of a ``$T$-coupled eddy flow''
	\begin{eqnarray*}
	\ve v_T = \frac {\mmean{T'\ve v'}} {\mean T}.
	\end{eqnarray*}
Since $\frac 1 2 \mean T^2$ is the eddy energy and is hence always positive, 
the eddy generation is therefore completely controlled by the divergence of
this flow. This simple formalism can be easily generalized to those with
momentum equations. 
In that case, the perfect transfer is 
a redistribution of kinetic energy between the mean and the 
eddy windows. The resulting transfer is referred to (\ref{eq:transfervec}).
For all the averaging schemes, it has the same form.

The formalism has been validated with a well-known barotropic instability
model, the Kuo model for the stability of a zonal atmospheric jet stream. 
Instability implies energy transfer from the background to perturbation,
or mean to eddy in this context.
We have seen a scenario of perfect transfer consistent with that inferred 
based on Kuo's theorem, while the traditional Reynolds stress extraction
does not agree with the inference.

An intuitive argument regarding the perfect transfer $\transfer$
is that the distribution of $\transfer$ is generally not in accordance 
to that of eddy energy or eddy energy growth, due to the presence of 
transport processes.
This has been testified in the wake control experiment.
In the context of turbulence, that is to say,
the rapid growth of turbulent energy does not necessarily correspond 
to turbulence production. It is not uncommon that, at a location
where perturbation is growing, the underlying process could be a transfer
in the inverse direction, i.e., a laminarization.
This argument has profound implication in real applications.
Turbulence control is such an example.
It suggests the optimal location to place a control be that of
positive $\transfer$, rather than that of turbulence growth, 
in order to take the advantage of the self-laminarization within
a turbulent flow. This conjecture has been testified in an exercise
of vortex shedding suppression with a cylinder wake.
By computation there are two lobes of strong positive $\transfer$ 
attaching to the cylinder on either side. 
We tried a surface-based suction on many places of the cylinder,
but the most effective places are those where the transfer processes 
within these two lobes are easiest to defeat. Other places are
not as effective as these two, in terms of energy saving.

The success of the wake suppression experiment implies the physically robust
quantity $\transfer$ may be useful for a variety of fluid control problems.
Specifically, it may come to help in selecting the location(s) to place
a passive control, or designing the performance functional
for an active control. The above experiment is an example for the former;
for the latter, we should be able to design some transfer-oriented
functional for the optimization. As we argued before, this should be
advantageous over those based on turbulence growth in light of
energy saving.

It should be pointed out that, in realistic flows, the signals are
generally not stationary, nor homogeneous, and as a result, the
Reynolds averaging cannot be replaced with an averaging over time or a
spatial dimension.
In such cases, the mean and eddy fields are not as simple as thus
reconstructed; the mean itself can be time varying. 
Besides, interactions may not be limited just between two windows.
A common process, mean-eddy-turbulence interaction, for example,
requires three distinct windows for a faithful representation.
All these difficulties will be overcome, and a new real problem-oriented
formalism will be realized in a forthcoming paper
after the introduction of a new analysis apparatus, multiscale
window transform, 
to replace the Reynolds averaging technique for a realistic 
mean-eddy-turbulence decomposition.

\begin{acknowledgments}

This work has been benefited from the important discussions with 
Allan Robinson, Howard Stone, Brian Farrell, and Glenn Flierl.
Joseph Pedlosky inspired the formalization of multiscale flux.
Part of the wake control experiments were run when the author visited
the Center for Turbulence Research at Stanford University and
NASA Ames Research Laboratory. Thanks are due to
Parviz Moin and Nagi Mansour for their kind invitation,
and Alan Wray for his generous help with the computing.
The author is particularly indebted to Meng Wang, who hosted the visit
and spent a lot of time discussing the issues raised in this work.

\end{acknowledgments}

\end{document}